\pdfoutput=1
\documentclass[11pt,twoside,a4paper,cmspaper,final,collab]{cms-tdr}
\def\svnVersion{02913f1}\def\svnDate{2023/07/25}\def\cmsCernNoTag{CERN-EP-2022-090}\def\cmsCernDate{\today}\def\cmsMessage{Published in Physical Review Letters as \href{http://dx.doi.org/10.1103/PhysRevLett.131.041803}{\doi{10.1103/PhysRevLett.131.041803}.}}
\begin{document}\cmsNoteHeader{B2G-22-003}

\newlength\cmsBigFigWidth
\newlength\cmsFigWidth
\ifthenelse{\boolean{cms@external}}{\setlength\cmsBigFigWidth{0.96\columnwidth}}{\setlength\cmsBigFigWidth{0.9\textwidth}}
\ifthenelse{\boolean{cms@external}}{\setlength\cmsFigWidth{0.96\columnwidth}}{\setlength\cmsFigWidth{0.66\textwidth}}
\newcommand{\mSD}{\ensuremath{m_\text{SD}}\xspace}
\newcommand{\mreg}{\ensuremath{m_\text{reg}}\xspace}
\newcommand{\mHH}{\ensuremath{m_{\PH\PH}}\xspace}
\newcommand{\Hbb}{\ensuremath{\PH\to\bbbar}\xspace}
\newcommand{\VVHH}{\ensuremath{{\PV\PV\PH\PH}}\xspace}
\newcommand{\HH}{\ensuremath{{\PH\PH}}\xspace}
\newcommand{\bbbb}{\ensuremath{\bbbar\bbbar}\xspace}
\newcommand{\kappalambda}{\ensuremath{\kappa_\lambda}\xspace}
\newcommand{\kappatop}{\ensuremath{\kappa_{\PQt}}\xspace}
\newcommand{\kappav}{\ensuremath{\kappa_\PV}\xspace}
\newcommand{\kappavv}{\ensuremath{\kappa_{2\PV}}\xspace}
\newcommand{\Dbb}{\ensuremath{D_{\bbbar}}\xspace}
\newcommand{\mjj}{\ensuremath{m_\text{jj}}\xspace}
\newcommand{\ggF}{\ensuremath{{\Pg\Pg\mathrm{F}}}\xspace}

\cmsNoteHeader{B2G-22-003}
\title{Search for nonresonant pair production of highly energetic Higgs bosons decaying to bottom quarks}

\date{\today}

\abstract{
   A search for nonresonant Higgs boson (\PH) pair production via gluon and vector boson (\PV) fusion is performed in the four-bottom-quark final state, using proton-proton collision data at 13\TeV corresponding to 138\fbinv collected by the CMS experiment at the LHC.
   The analysis targets Lorentz-boosted \PH pairs identified using a graph neural network.
   It constrains the strengths relative to the standard model of the \PH self-coupling and the quartic \VVHH couplings, $\kappavv$, excluding $\kappavv = 0$ for the first time, with a significance of 6.3 standard deviations when other \PH couplings are fixed to their standard model values.
}

\hypersetup{
   pdfauthor={CMS Collaboration},
   pdftitle={Search for nonresonant pair production of highly energetic Higgs bosons decaying to bottom quarks},
   pdfsubject={CMS},
   pdfkeywords={HH, machine learning, GNN, Higgs boson, bottom quarks, ParticleNet}
}

\maketitle

The discovery of a Higgs boson by the ATLAS and CMS Collaborations at the CERN LHC in 2012~\cite{Aad:2012tfa,Chatrchyan:2012ufa,CMS:2013btf} started a new era in experimental high energy physics.
Since its original observation, significant efforts have been made to measure the properties of the Higgs boson, including its self-coupling.
The production of a pair of Higgs bosons (\HH) is a rare process that provides access to direct measurements of the Higgs boson trilinear self-coupling ($\lambda$) and the quartic couplings between two Higgs bosons and two vector bosons.

The \HH production takes place primarily via gluon-gluon fusion (\ggF) and vector boson fusion (VBF).
In the standard model (SM), the \ggF cross section at $\sqrt{s} = 13\TeV$ for a Higgs boson mass of 125\GeV, calculated at next-to-next-to-leading order (NNLO) precision in perturbative quantum chromodynamics (QCD), is $31.1^{+2.1}_{-7.2}\unit{fb}$~\cite{Grazzini:2018bsd,Baglio:2020wgt} and is sensitive to the values of $\lambda$ and the top quark Yukawa coupling, $y_{\PQt}$.
Variations from their SM values are parametrized as $\kappalambda = \lambda / \lambda^\text{SM}$ and $\kappatop = y_{\PQt} / y_{\PQt}^\text{SM}$.
The latter has been measured to be consistent with the SM~\cite{CMS:2018uag,ATLAS:2019nkf}.
The VBF cross section, calculated at next-to-NNLO (N$^3$LO) precision, is $1.726 \pm 0.036\unit{fb}$~\cite{Dreyer:2018qbw} and depends on the strength of the self-coupling and on the interaction of a pair of \PW or \PZ bosons (jointly denoted as \PV) with a single Higgs boson ($\PV\PV\PH$) or a pair of Higgs bosons ($\PV\PV\HH$), whose values with respect to the SM predictions are parametrized as \kappav and \kappavv, respectively.
The ATLAS and CMS Collaborations have performed studies of \HH production at $\sqrt{s} = 7$, 8, and 13\TeV in the separate $\bbbar\gamma\gamma$~\cite{Aad:2014yja,Sirunyan:2018iwt,CMS:2020tkr,ATLAS:2021ifb}, $\bbbar\TT$~\cite{Sirunyan:2017tqo,Aaboud:2018sfw}, $\bbbb$~\cite{Aad:2015uka,Aaboud:2016xco,Aaboud:2018knk,Sirunyan:2018tki,Aad:2020kub,CMS:2022cpr}, and $\bbbar\PV\PV$~\cite{,Sirunyan:2017guj,Aaboud:2018zhh,Aad:2019yxi,Sirunyan:2020qcq} channels, as well as in combinations of channels~\cite{Aad:2015xja,Sirunyan:2018ayu,Aad:2019uzh}.

This Letter reports the results of a search for nonresonant \HH production via both the \ggF and VBF modes in the channel where each Higgs boson decays to a bottom quark-antiquark pair (\bbbar).
It is based on data from proton-proton collisions at the LHC at $\sqrt{s} = 13\TeV$, collected by the CMS experiment in 2016--2018, corresponding to an integrated luminosity of 138\fbinv.
We select events with both Higgs bosons in the highly Lorentz-boosted regime, \ie, with sufficiently large transverse momentum (\pt) for the decay products of each Higgs boson to become merged into a single large-radius jet.
Events are separated into mutually exclusive categories targeting either \ggF or VBF \HH production, with the latter requiring a characteristic VBF signature of two additional small-radius jets in opposite forward regions.
The \ggF categories are designed to select SM \ggF \HH signal events, whereas the VBF categories target non-SM coupling hypotheses that can dramatically increase the VBF \HH cross section, especially for highly Lorentz-boosted Higgs boson pairs~\cite{Dolan:2013rja,Dolan:2015zja,Bishara:2016kjn,Arganda:2018ftn}.
Such non-SM couplings can arise for example in the models where the Higgs boson is a pseudo-Goldstone boson from a new {\TeVns}-scale strong interaction~\cite{Bishara:2016kjn,Giudice:2007fh}, and a case of particular interest is $\kappavv = 0$, corresponding to vanishing \VVHH couplings.
The two dominant background processes, namely QCD multijet production, consisting uniquely of jets produced through the strong interaction, and top quark-antiquark pair (\ttbar) production, are estimated utilizing dedicated data samples enriched in these processes.
One of the main challenges for this search is the efficient reconstruction of the decay of a pair of Higgs bosons, while rejecting jets originating from light-flavor quarks and gluons.
For this purpose, we use a novel graph neural network (GNN) algorithm, ParticleNet~\cite{Qu:2019gqs}, to select large-radius jets resulting from \Hbb decays, increasing signal purity in a mass-independent way.
The search complements that in Ref.~\cite{CMS:2022cpr}, which targets the same \bbbb final state in the non-boosted regime and covers a region of signal phase space that is largely disjoint from that of the present analysis.

The CMS apparatus~\cite{CMS:2008xjf} is a multipurpose, nearly hermetic detector, designed to trigger on~\cite{CMS:2016ngn,CMS:2020cmk} and identify electrons, muons, photons, and charged and neutral hadrons~\cite{CMS:2020uim,CMS:2018rym,CMS:2014pgm}.
A global reconstruction algorithm~\cite{CMS:2017yfk} combines the information provided by the all-silicon inner tracker and by the crystal electromagnetic (ECAL) and brass-scintillator hadron calorimeters, operating inside a 3.8\unit{T} superconducting solenoid, with data from gas-ionization muon detectors interleaved with the solenoid return yoke~\cite{CMS:2018jrd,CMS:2016lmd,CMS:2019ctu}.
The primary vertex (PV) is taken to be the vertex corresponding to the hardest scattering in the event, evaluated using tracking information alone~\cite{CMS-TDR-15-02}.

Simulated samples for the \ggF process are generated at next-to-leading order (NLO) precision using \POWHEG~2.0~\cite{POWHEG1,POWHEG2,POWHEG3,Bagnaschi:2011tu,Heinrich:2017kxx,Jones:2017giv,Buchalla:2018yce,Heinrich:2019bkc,Heinrich:2020ckp}, and corrected as a function of \HH mass based on Ref.~\cite{Davies:2019dfy}.
The VBF \HH samples are generated at leading order (LO) precision using \MGvATNLO 2.6.5~\cite{Alwall:2014hca}.
A range of samples corresponding to different combinations of the \kappav, \kappavv, and \kappalambda coupling modifiers is generated.
Samples for other coupling combinations are constructed as linear combinations of the original generated samples by applying appropriate event weights as described in Ref.~\cite{CMS:2022cpr}.
The \ggF samples are normalized to the NNLO cross sections~\cite{Dawson:1998py,Shao:2013bz,deFlorian:2013jea,deFlorian:2015moa,Borowka:2016ehy,Grazzini:2018bsd,Baglio:2018lrj,Baglio:2020wgt} corresponding to the coupling values considered.
The VBF sample with SM couplings is normalized to the N$^3$LO cross section~\cite{Dreyer:2018qbw}, and the same N$^3$LO/LO correction is applied to the VBF samples with modified couplings.

Samples for the \ttbar background process are generated at NLO precision using \POWHEG~v2.0~\cite{POWHEG1,POWHEG2,POWHEG3,Frixione:2007nw,POWHEG-TT} with the FxFx jet matching and merging scheme~\cite{Frederix:2012ps}, and normalized to the theoretical cross section calculated at NNLO precision using {{\textsc{Top++} v2.0}}~\cite{Czakon:2011xx}.
The differential \ttbar cross section as a function of the top quark \pt is corrected to the NNLO QCD + NLO electroweak precision~\cite{Czakon:2017wor}.
Samples for other background processes, such as the QCD multijet, single top quark, \PW and \PZ single-boson and diboson production processes, as well as single Higgs boson production in all relevant production modes, are also generated.
Parton showering, hadronization, and the underlying event are modeled by \PYTHIA8.205~\cite{Sjostrand:2014zea} with parameters set by the CUETP8M1~\cite{Khachatryan:2015pea} and CP5 tunes~\cite{CMS:2019csb} used for samples simulating the 2016 and 2017--2018 conditions, respectively.
The NNPDF~3.0~\cite{Ball:2014uwa} and 3.1~\cite{Ball:2017} parton distribution functions (PDFs) are used in the generation of all simulated samples.
The \GEANTfour~\cite{GEANT4} package is used to model the response of the CMS detector, and simulated minimum bias interactions are mixed with the hard interactions in simulated events to model additional proton-proton interactions within the same or nearby bunch crossings (pileup).
The simulated events are weighted to match the pileup distributions measured in data.

The reconstructed particles are clustered into jets using the anti-\kt algorithm~\cite{Cacciari:2008gp} implemented in the \FASTJET package~\cite{Cacciari:2011ma} with a distance parameter of 0.4 (small-radius jets) or 0.8 (large-radius jets).
To mitigate the effect of pileup on small-radius jets, the charged-hadron subtraction algorithm~\cite{CMS:2017yfk} is applied.
For large-radius jets, the pileup-per-particle identification algorithm~\cite{Bertolini:2014bba,Sirunyan:2020foa} is used to assign a weight to each particle prior to jet clustering based on the likelihood of the particle originating from the PV.
Further corrections are applied to the jet energy as a function of the jet pseudorapidity ($\eta$) and \pt to account for detector response nonlinearities~\cite{CMS:2016lmd}.
The missing transverse momentum vector \ptvecmiss is computed as the negative vector sum of the transverse momenta of all the reconstructed particles in an event, and its magnitude is denoted as \ptmiss~\cite{CMS:2019ctu}.
The \ptvecmiss is modified to account for corrections to the energy scale of the reconstructed jets in the event.

Electrons are reconstructed and their momentum is estimated by combining the momentum measurement from the tracker, the energy of the corresponding ECAL cluster, and the energy sum of all bremsstrahlung photons spatially compatible with originating from the electron track~\cite{CMS:2020uim}.
Muons are reconstructed as tracks in the central tracker, consistent with either a track or several hits in the muon chambers, and their momentum is obtained from the curvature of the corresponding track~\cite{CMS:2018rym}.
Electron candidates are reconstructed within the tracker acceptance of $\abs{\eta} < 2.5$, whereas muon candidates are required to be within the muon chamber acceptance of $\abs{\eta} < 2.4$.
The electron and muon candidates are required to have $\pt > 10\GeV$.
Only tracks consistent with the PV are associated with the electrons or muons, and additional identification criteria~\cite{CMS:2015xaf,CMS:2018rym} are applied to improve the purity of the lepton selection.
Isolation criteria~\cite{CMS:2016krz} are used to distinguish leptons originating at the PV from those originating from bottom or charm hadron decays.

Events are required to contain at least two large-radius jets, and the two with the highest \pt are considered as Higgs boson candidates.
A combination of several trigger algorithms is used, all requiring the total hadronic transverse energy in the event ($\HT$) or jet $\pt$ to be above a given threshold.
After removing the remnants of soft radiation~\cite{Krohn:2009th}, a minimum triggering threshold on the jet mass is imposed in order to reduce the \HT or \pt thresholds relative to a trigger algorithm with no jet mass requirement but of an equivalent rate.
The trigger efficiency varies between 10 and 95\% for jets with \pt in the range from 300 to 450\GeV and is fully efficient for jets with $\pt>500\GeV$.

The GNN algorithms~\cite{neuraljets,Qu:2019gqs,Moreno:2019bmu,Moreno:2019neq} have been shown to achieve state-of-the-art performance in jet classification~\cite{Kasieczka:2019dbj}, and this analysis is among the first to use the ParticleNet GNN algorithm~\cite{Qu:2019gqs} to discriminate between \Hbb and QCD-induced jets.
Its inputs comprise 20 measured properties of jet constituent particles, such as the \pt, the angular separation between the particle and the jet axis, and the transverse and longitudinal impact parameters relative to the PV.
The inputs also include eleven properties of the secondary vertices~\cite{CMS:2017wtu} within the jet cone, such as the displacement, \pt, mass, and the number of associated tracks.
Given the relatively large masses, long lifetimes, and high-\pt and displaced decay products of \PQb hadrons, this low-level tracking and vertexing information is important for identifying jets containing \PQb hadron decays~\cite{CMS:2012feb}.
The algorithm treats each jet as an unordered set of its constituents, considers their correlations at different scales to extract jet features, and returns an output corresponding to the probability that the jet belongs to either the \Hbb or the QCD multijet class.
As shown in Ref.~\cite{Sirunyan:2020lcu}, algorithms like this one that exploit the correlations between many low-level input features outperform those that are based on fewer high-level discriminating variables for \Hbb identification.
To achieve independence from the jet mass, the classifier is trained using dedicated simulated samples containing spin-0 particles with a uniform mass spectrum between 15 and 250\GeV and decaying to quark-antiquark pairs, as well as a QCD multijet sample~\cite{CMS-DP-2020-002}.
In addition, jets in each sample are reweighted to obtain uniform distributions in both \pt and mass for the training.
The ParticleNet algorithm yields significant improvements both in terms of performance and jet mass decorrelation compared to previous jet classifier algorithms~\cite{CMS-DP-2020-002,Sirunyan:2020lcu}.
The discriminant (\Dbb) is calibrated using data and simulated samples dominated by QCD multijet events~\cite{CMS-DP-2022-005}.
A boosted decision tree (BDT) classifier is used to select a sample of jets enriched in gluon splittings to \bbbar that yields a \Dbb distribution similar to the \HH signal.
A fit to data is performed to extract \pt-dependent simulation-to-data correction factors, typically ranging from 0.9 to 1.2.
They are applied to correct large-radius jet selection efficiencies in the signal events.
Three working points (WPs) based on \Dbb are used: these tight, medium, and loose WPs correspond to \Hbb selection efficiencies of approximately 60, 80, and $90\%$, and QCD background misidentification rates of approximately 0.3, 1, and $2\%$, respectively.

The search for Higgs boson pairs relies on the ability to reconstruct accurately the masses of the two Higgs boson candidates.
The soft-drop (SD) algorithm~\cite{Larkoski:2014wba} with angular exponent $\beta=0$ and soft radiation fraction $z=0.1$, also known as the modified mass-drop algorithm~\cite{Dasgupta:2013ihk}, is usually applied to the Higgs boson jet candidate to remove soft and wide-angle radiation.
The SD algorithm occasionally over-subtracts genuine jet constituents that account for a large fraction of the jet momentum, resulting in a jet mass (\mSD) that is negligible even for large-mass resonances.
To improve the jet mass estimation, we introduce a regression algorithm based on the ParticleNet GNN architecture to predict the jet mass~\cite{CMS-DP-2021-017}.
To train the algorithm, the target mass value that the algorithm aims to predict is chosen as the true resonance mass for the heavy spin-0 particles, or the SD mass of particle-level jets for QCD multijet events.
Particle-level jets are defined by clustering stable simulated final-state particles, excluding neutrinos, using the same procedure as for reconstructed jets, and they are geometrically matched to the reconstructed jets.
The training is performed with the same samples and a similar event weighting scheme as for jet classification.
Data control samples, dominated by \ttbar events, are used to calibrate the jet mass from the regression (\mreg).
The small corrections to mass scale ($<$1\%) and resolution ($\approx$3\%) are then applied to all simulated samples in the rest of the analysis, accompanied by systematic uncertainties that account for the differences in regression performance between the jets originating from \PW and Higgs boson decays.

Events containing at least two large-radius jets with $\pt>300\GeV$ and $\abs{\eta} < 2.4$, and no isolated electrons or muons, are selected and grouped into either the \ggF or the VBF categories.

For the VBF categories, we require the two large-radius jets to have $\mreg$ between 50 and 200\GeV.
The search region (SR) is defined as the subset of events with \mreg of the \pt-leading (\pt-subleading) jet in the range 110--150 (90--145)\GeV, while the remaining events are used as part of the background estimation method described below.
As the VBF process is characterized by the presence of two forward jets with a large dijet invariant mass and a gap in $\eta$, we also require two additional small-radius jets, referred to as VBF jet candidates, with $\pt > 25\GeV$ and $\abs{\eta}<4.7$.
They are required to be separated from both Higgs boson candidate jets by a minimal distance of 1.2 in the $\eta$-$\phi$ plane, where $\phi$ denotes the azimuthal angle.
If more than two such jets are found, the two with the highest \pt are selected.
The absolute difference in pseudorapidity ($\Delta\eta^\text{VBF}_\text{jj}$) between the two VBF jets must be larger than 4.0, and the invariant mass of the two VBF jets ($\mjj^\text{VBF}$) must be larger than 500\GeV.
To increase the sensitivity for the highly boosted Higgs boson pairs from VBF production with non-SM couplings, the $\pt$-leading ($\pt$-subleading) large-radius jet is required to have $\pt > 500$ (400)\GeV.
Additionally, the azimuthal angle between the two large-radius jets ($\Delta\phi_{\text{j}_{1}\text{j}_{2}}$) is required to be larger than 2.6, and the absolute difference in pseudorapidity $\Delta\eta_{\text{j}_{1}\text{j}_{2}}$ must be less than 2.0.
Three VBF event categories are defined based on the \Dbb scores of the two Higgs boson candidate jets.
Events in the high-purity (HP) category must have both jets passing the tight \Dbb WP; events not in the HP category are categorized as medium purity (MP) if both jets pass the medium \Dbb WP; and events not in the HP or MP categories are categorized as low purity (LP) if both jets pass the loose \Dbb WP.
The definitions of the three categories based on the corresponding \Dbb WPs were optimized to provide the best expected significance for the VBF \HH signal hypotheses of interest such as $\kappavv = 0$ (with other couplings at their SM values), while enabling a robust background estimation and reliable \Dbb calibration in all categories.

Events not assigned to any of the VBF categories are considered for the \ggF categories.
The two selected large-radius jets are ordered by the value of \Dbb.
The \Dbb-leading jet is required to have $\mSD > 50\GeV$, while the \Dbb-subleading jet is required to have $\mreg > 50\GeV$.
A BDT is trained to discriminate between the \HH signal and QCD multijet or \ttbar background processes.
Input variables to the BDT include: the \pt and substructure variable $\tau_{32}$, used to identify three-prong jets like those arising from fully hadronic top quark decays~\cite{Thaler:2010tr,Sirunyan:2020lcu}, of each of the two selected large-radius jets; the \mSD, $\eta$, and \Dbb of the \Dbb-leading jet; the \ptmiss; the \pt, $\eta$, and mass of the dijet system (\mHH); and the ratios $\pt{}_{\text{j}_1}/\mHH$, $\pt{}_{\text{j}_2}/\mHH$, and $\pt{}_{\text{j}_2}/\pt{}_{\text{j}_1}$.
The \mSD is used as an input to the BDT instead of \mreg in order to avoid correlations between the mass regression and the BDT.
The simulation modeling of the BDT input variables is validated in signal and background control regions and shown to be accurate.
Based on the BDT output score, tight, medium, and loose WPs are defined, corresponding to \HH signal selection efficiencies of approximately 23, 27 and 33\%, and QCD and \ttbar background misidentification rates of approximately 1, 2 and 12\%, respectively.
Three SR event categories targeting \ggF production are constructed.
Events in the highest signal purity category 1 are required to pass the tight BDT WP and the \Dbb-subleading large-radius jet is required pass the tight \Dbb WP.
Events not in category 1 are grouped into category 2 if they pass the tight BDT WP and the \Dbb-subleading jet passes the medium \Dbb WP, or if they fail the tight but pass the medium BDT WP and the \Dbb-subleading jet passes the tight \Dbb WP.
Finally, events not in categories 1 or 2 are grouped into category 3 if they pass the loose BDT WP and the \Dbb-subleading jet passes the medium \Dbb WP.
The categorization thresholds are chosen to maximize the expected sensitivity to the SM \ggF \HH signal.

After all selection requirements, the remaining SM background mainly consists of two processes: \ttbar and QCD multijet production.
These background contributions are estimated by fitting the corresponding distributions simultaneously via a maximum likelihood fit to data as detailed in the following, similarly to previous CMS searches for boosted Higgs boson signals~\cite{Sirunyan:2017isc,Sirunyan:2018qca,Sirunyan:2020hwz}.
Other backgrounds include single Higgs boson production in the $\ttbar\PH$ and $\PV\PH$ modes, {\PV}+jets production, and diboson production.
They are estimated from simulation in the \ggF event categories, and found to be negligible in the VBF event categories.

A control region (CR) enriched in QCD multijet events is selected by changing the requirement on the \Dbb discriminant.
For the VBF categories, this CR is defined by both Higgs boson candidate jets failing the loose \Dbb WP.
Since the \Dbb discriminant is designed and measured to be independent of the jet mass, the sidebands of the mass distribution of the \pt-subleading large-radius jet, namely $\mreg\in[50, 100]\cup [145, 200]\GeV$, are used to obtain transfer factors that account for the ratio of selection efficiencies in the QCD-enriched low-\Dbb CR and the high-\Dbb VBF SRs.
A separate transfer factor is derived using the \pt-subleading large-radius jet mass sidebands for each \mHH interval in each search category (HP, MP, and LP).
For the \ggF categories, the QCD-enriched CR is defined by the \Dbb-subleading large-radius jet failing the medium \Dbb WP and passing the loose BDT WP.
The QCD multijet background in the \ggF SRs is estimated using the \mreg distribution of the \Dbb-subleading jet, assuming a constant transfer factor~\cite{Sirunyan:2020hwz}.
We have checked whether a higher-order polynomial dependence on the mass shape is necessary using a Fisher $F$-test~\cite{ref:ftest}, and Kolmogorov--Smirnov~\cite{KS} and saturated model~\cite{Baker:1983tu,lindsey1996parametric} goodness-of-fit tests, but a constant factor is found to be sufficient.

For the VBF categories, an auxiliary sample enriched in \ttbar events containing one leptonically decaying \PW boson is used to extract two corrections for the \ttbar background estimate: one for the difference in \Hbb misidentification efficiency in \ttbar events between data and simulation, measured separately for each \Dbb WP, and the other for the overall normalization of the \ttbar process, measured inclusively.
We define this sample with a set of selections closely following the definition of the \ttbar-enriched region in Ref.~\cite{Sirunyan:2020lcu}.
For the \ggF categories, the simulation predictions of the BDT and \mreg distributions for the top quark background are validated and corrected using a similar \ttbar sample containing one leptonic \PW boson decay, with $\tau_{32} < 0.46$ required for two large-radius jets.

The dominant uncertainty in the analysis, corresponding to 62\% relative to the extracted signal yield is the statistical uncertainty due to limited size of the data sample.
Other sources of uncertainty include: the QCD multijet background estimate (30\%); the simulation modeling of the \Dbb shape and selection efficiency (15\%); the jet energy and mass scale and resolution (13\%); the theoretical uncertainties from the PDFs, missing higher-order QCD corrections, and initial- and final-state radiation ($11\%$); the modeling of the top quark background ($7\%$); the trigger efficiency measurement ($4\%$); the integrated luminosity measurement~\cite{CMS-LUM-17-003,CMS-PAS-LUM-17-004,CMS-PAS-LUM-18-002} ($2\%$); and the pileup modeling ($2\%$).
The impact of each uncertainty is evaluated by computing the uncertainty excluding that source and subtracting it in quadrature from the total uncertainty.

\begin{figure}[htb!]
   \centering
   \includegraphics[width=\cmsFigWidth]{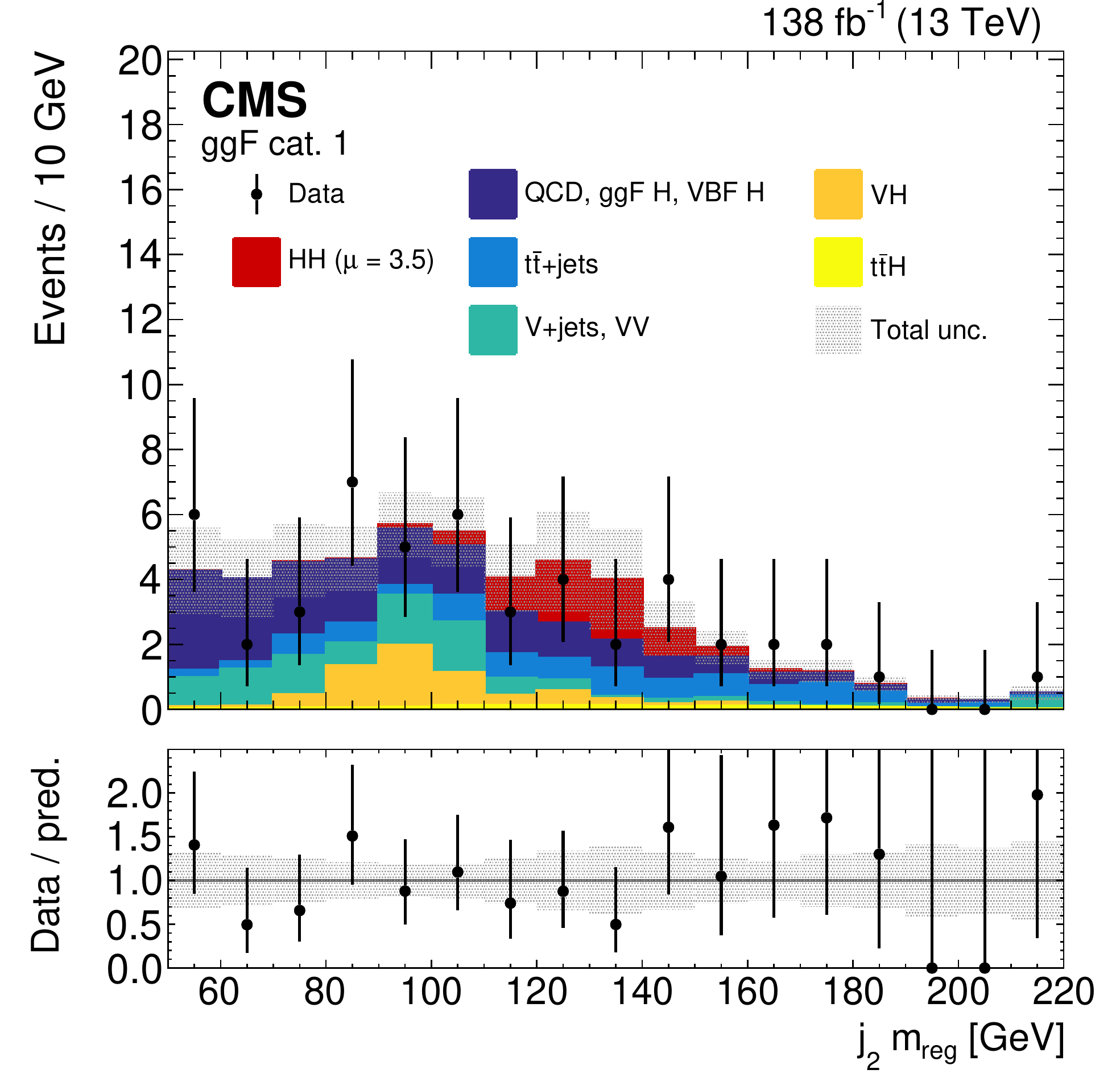}
   \caption{The data and fitted signal and background distributions for the \Dbb-subleading jet regressed mass are shown for the \ggF BDT event category 1, the category accounting for most of the sensitivity to the \ggF \HH signal.
      The SM \HH ($\kappavv=\kappav=\kappalambda=1$) signal is shown scaled to the best fit signal strength $\mu=3.5$.
      The lower panel shows the ratio of the data and the total prediction, with its uncertainty represented by the shaded band.
      The error bars on the data points represent the statistical uncertainties.
      \label{fig:ggHH_fitresult}}
\end{figure}

\begin{figure}[thb!]
   \centering
   \includegraphics[width=\cmsFigWidth]{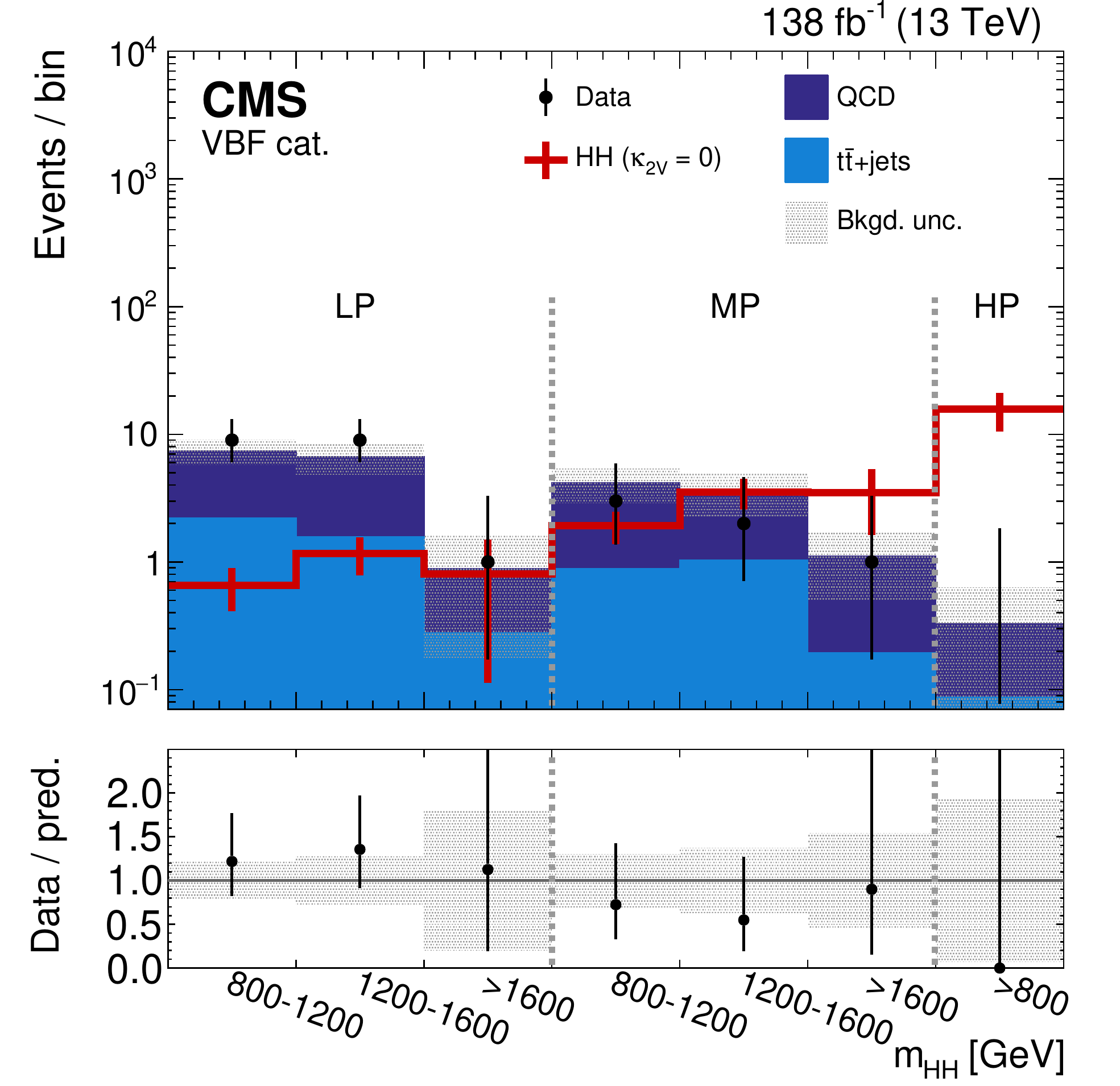}
   \caption{The distributions of the invariant mass of the \HH system after a background-only fit to the data, for the VBF low-purity (LP), medium-purity (MP), and high-purity (HP) categories.
      The VBF signal for $\kappavv=0$, $\kappav=\kappalambda=1$, is shown in red with the vertical error bar indicating the prefit uncertainty.
      The lower panel shows the ratio of the data and the total background prediction, with its uncertainty represented by the shaded band.
      The error bars on the data points represent the statistical uncertainties.
      \label{fig:VBF_fitresult}}
\end{figure}

A binned maximum likelihood fit to the observed \mHH, \Dbb, \mreg, and BDT distributions is performed using the sum of the signal and background contributions.
Because part of the VBF signal can enter the \ggF categories, \eg, caused by failing to reconstruct the VBF jets or the Higgs boson candidate jet \pt being too low, the fit is performed simultaneously in all \ggF and VBF categories, including the CRs.
The results of the fit in the \ggF BDT event category 1, which dominates the overall sensitivity, are shown in Fig.~\ref{fig:ggHH_fitresult}.
The results of the fit in the VBF LP, MP, and HP categories are shown in Fig.~\ref{fig:VBF_fitresult}.

\begin{figure}[htb!]
   \centering
   \includegraphics[width=\cmsBigFigWidth]{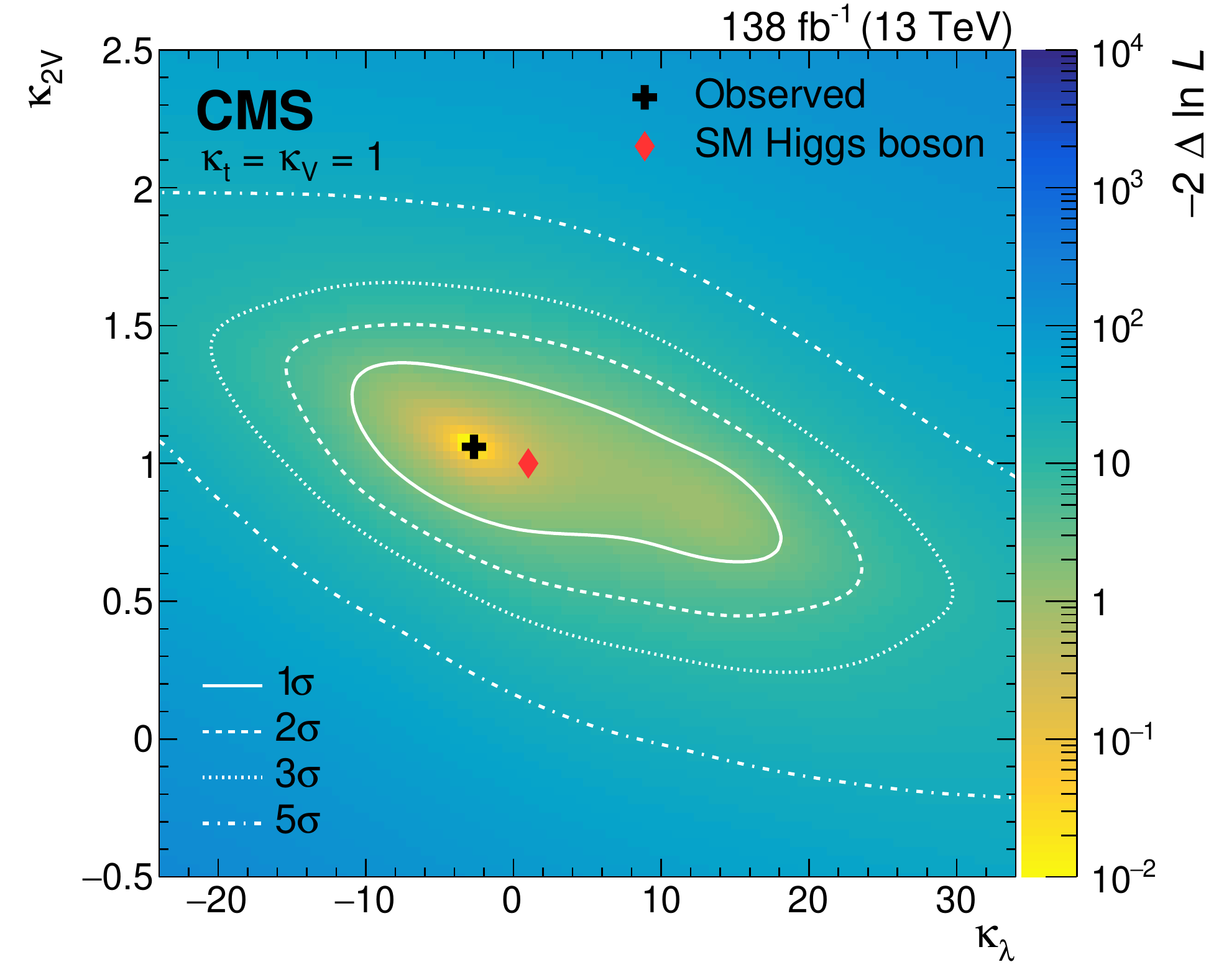}
   \caption{Two-parameter profile likelihood test statistic ($-2\Delta\ln\mathcal{L}$) scan in data as a function of \kappalambda and \kappavv.
      The black cross indicates the minimum, while the red diamond marks the SM expectation ($\kappalambda = \kappavv = 1$).
      The gray solid, dashed, dotted, and dash-dotted contours enclose the 1, 2, 3, and 5\,$\sigma$ \CL regions, respectively.
      \label{fig:2d}}
\end{figure}

The test statistic chosen to determine the signal yield is based on the profile likelihood ratio~\cite{LHCCLs}.
Systematic uncertainties are incorporated into the analysis via nuisance parameters and treated according to the frequentist paradigm.
To ensure that the background estimation methods and the resulting statistical model are able to disentangle possible signal from background, we perform signal injection tests.
We generate various pseudo-experiment data sets corresponding to different signal hypotheses, fit these data sets, and verify that the extracted signal parameters correspond to the injected ones within the fitted uncertainties.
Then we proceed to fit the observed data, for which the signal strength modifier, defined as the ratio of the measured to the expected SM \HH production, extracted from a combined fit of all categories is found to be $\mu=3.5^{+3.3}_{-2.5}$.
The best fit value of each parameter of interest and approximate 1, 2, 3, and 5 standard deviation ($\sigma$) confidence level (\CL) intervals are extracted following the procedure described in Section~3.2 of Ref.~\cite{Khachatryan:2014jba}.
Figure~\ref{fig:2d} shows the profile likelihood test statistic scan in data over the \kappalambda-\kappavv plane.

Upper limits on the \HH production cross section at 95\% \CL based on the \CLs criterion~\cite{CLS2,CLS1} are obtained using asymptotic formulas~\cite{Cowan:2010js}.
The upper limit on the total \HH production cross section is observed (expected) to be 9.9 (5.1) relative to the SM prediction.
The observed (expected) 95\% \CL interval for \kappalambda is found to be $[-9.9, 16.9]$ ($[-5.1, 12.2]$), assuming all other Higgs boson couplings to be at their SM values.
Similarly, the observed (expected) 95\% \CL interval for \kappavv is found to be $[0.62, 1.41]$ ($[0.66, 1.37]$).
The observed (expected) interval for \kappav is found to be $[-1.18, -0.78] \cup [0.79, 1.19]$ ($[-1.15, -0.80] \cup [0.82, 1.21]$).
The lack of a large observable VBF \HH signal establishes a nonzero value of \kappavv for the first time with a significance of 6.3\,$\sigma$,  when other Higgs boson couplings are fixed to their SM values.
Tabulated results are provided in the HEPData record for this analysis~\cite{hepdata}.

In summary, a search for nonresonant Higgs boson pair (\HH) production via gluon-gluon fusion and vector boson (\PV) fusion in the final state with two bottom quark-antiquark ($\bbbar$) pairs has been presented.
The search is focused on the phase space region where both Higgs bosons are highly Lorentz boosted so that each Higgs boson is reconstructed as a large-radius jet.
A novel algorithm based on a graph neural network is applied to identify the jets that correspond to \Hbb decays.
The data are found to agree with the background-only hypothesis, and an observed (expected) upper limit at 95\% confidence level is set to 9.9 (5.1) relative to the standard model cross section.
This represents a factor of 30 improvement over the previous best search for a pair of boosted \Hbb jets, which used only data collected in 2016 (36\fbinv) and less advanced methods for \Hbb identification and event selection~\cite{Sirunyan:2018qca}.
Upper limits on the production cross section are set as a function of the coupling modifier parameters \kappalambda and \kappavv, which parametrize the strengths of the Higgs boson self-coupling, and the quartic \VVHH couplings, respectively, relative to their standard model values.
The values of \kappalambda and \kappavv are observed (expected) to be in the ranges $[-9.9, 16.9]$ ($[-5.1, 12.2]$) and $[0.62, 1.41]$ ($[0.66, 1.37]$), respectively, excluding $\kappavv = 0$ for the first time, with a significance of 6.3 standard deviations, when all the Higgs boson couplings, except for the one being scanned, are set to their standard model values.

\begin{acknowledgments}
   We congratulate our colleagues in the CERN accelerator departments for the excellent performance of the LHC and thank the technical and administrative staffs at CERN and at other CMS institutes for their contributions to the success of the CMS effort.
   In addition, we gratefully acknowledge the computing centers and personnel of the Worldwide LHC Computing Grid and other centers for delivering so effectively the computing infrastructure essential to our analyses.
   Finally, we acknowledge the enduring support for the construction and operation of the LHC, the CMS detector, and the supporting computing infrastructure provided by the following funding agencies: BMBWF and FWF (Austria); FNRS and FWO (Belgium); CNPq, CAPES, FAPERJ, FAPERGS, and FAPESP (Brazil); MES and BNSF (Bulgaria); CERN; CAS, MoST, and NSFC (China); MINCIENCIAS (Colombia); MSES and CSF (Croatia); RIF (Cyprus); SENESCYT (Ecuador); MoER, ERC PUT and ERDF (Estonia); Academy of Finland, MEC, and HIP (Finland); CEA and CNRS/IN2P3 (France); BMBF, DFG, and HGF (Germany); GSRI (Greece); NKFIH (Hungary); DAE and DST (India); IPM (Iran); SFI (Ireland); INFN (Italy); MSIP and NRF (Republic of Korea); MES (Latvia); LAS (Lithuania); MOE and UM (Malaysia); BUAP, CINVESTAV, CONACYT, LNS, SEP, and UASLP-FAI (Mexico); MOS (Montenegro); MBIE (New Zealand); PAEC (Pakistan); MES and NSC (Poland); FCT (Portugal); MESTD (Serbia); MCIN/AEI and PCTI (Spain); MOSTR (Sri Lanka); Swiss Funding Agencies (Switzerland); MST (Taipei); MHESI and NSTDA (Thailand); TUBITAK and TENMAK (Turkey); NASU (Ukraine); STFC (United Kingdom); DOE and NSF (USA).
\end{acknowledgments}
\bibliography{auto_generated}
\cleardoublepage \appendix\section{The CMS Collaboration \label{app:collab}}\begin{sloppypar}\hyphenpenalty=5000\widowpenalty=500\clubpenalty=5000
\cmsinstitute{Yerevan Physics Institute, Yerevan, Armenia}
{\tolerance=6000
A.~Tumasyan\cmsAuthorMark{1}\cmsorcid{0009-0000-0684-6742}
\par}
\cmsinstitute{Institut f\"{u}r Hochenergiephysik, Vienna, Austria}
{\tolerance=6000
W.~Adam\cmsorcid{0000-0001-9099-4341}, J.W.~Andrejkovic, T.~Bergauer\cmsorcid{0000-0002-5786-0293}, S.~Chatterjee\cmsorcid{0000-0003-2660-0349}, K.~Damanakis\cmsorcid{0000-0001-5389-2872}, M.~Dragicevic\cmsorcid{0000-0003-1967-6783}, A.~Escalante~Del~Valle\cmsorcid{0000-0002-9702-6359}, P.S.~Hussain\cmsorcid{0000-0002-4825-5278}, M.~Jeitler\cmsAuthorMark{2}\cmsorcid{0000-0002-5141-9560}, N.~Krammer\cmsorcid{0000-0002-0548-0985}, L.~Lechner\cmsorcid{0000-0002-3065-1141}, D.~Liko\cmsorcid{0000-0002-3380-473X}, I.~Mikulec\cmsorcid{0000-0003-0385-2746}, P.~Paulitsch, F.M.~Pitters, J.~Schieck\cmsAuthorMark{2}\cmsorcid{0000-0002-1058-8093}, R.~Sch\"{o}fbeck\cmsorcid{0000-0002-2332-8784}, D.~Schwarz\cmsorcid{0000-0002-3821-7331}, S.~Templ\cmsorcid{0000-0003-3137-5692}, W.~Waltenberger\cmsorcid{0000-0002-6215-7228}, C.-E.~Wulz\cmsAuthorMark{2}\cmsorcid{0000-0001-9226-5812}
% ...
\par}
\vskip\cmsinstskip
\dag:~Deceased\\
$^{1}$Also at Yerevan State University, Yerevan, Armenia\\
$^{2}$Also at TU Wien, Vienna, Austria\\
% ...
\end{sloppypar}
\end{document}